\documentclass[twocolumn,amsmath,amssymb,prl,superscriptaddress,longbibliography]{revtex4-1}

\setcitestyle{super}
\usepackage{bm}

\usepackage[final,letterspace=-40]{microtype}
\usepackage[colorlinks=true, linkcolor=blue, urlcolor=blue,  citecolor=blue, anchorcolor=blue]{hyperref}

\usepackage{mathpazo} 

\usepackage{graphicx}
\usepackage{dcolumn}
\usepackage{bm}
\usepackage{helvet}

\usepackage{caption}
\DeclareCaptionLabelFormat{plain}{Figure #2 $\boldsymbol |$ }
\begin{document}

\title{Classical entanglement underpins the propagation invariance of space-time wave packets}

\author{H. Esat Kondakci}
\email{esat@knights.ucf.edu}
\affiliation{CREOL, The College of Optics \& Photonics, University of Central Florida, Orlando, Florida 32816, USA}
\author{Miguel A. Alonso}
\affiliation{The Institute of Optics, University of Rochester, Rochester, NY 14627, USA}
\affiliation{Aix Marseille Univ., CNRS, Centrale Marseille, Institut Fresnel, UMR 7249, 13397 Marseille Cedex 20, France}
\author{Ayman F. Abouraddy}
\email{raddy@creol.ucf.edu}
\affiliation{CREOL, The College of Optics \& Photonics, University of Central Florida, Orlando, Florida 32816, USA}

\begin{abstract} 
\noindent
Space-time wave packets are propagation-invariant pulsed beams that travel in free space without diffraction or dispersion by virtue of tight correlations introduced into their spatio-temporal spectrum. Such correlations constitute an embodiment of classical entanglement between continuous degrees of freedom. Using a measure of classical entanglement based on the Schmidt number of the field, we demonstrate theoretically and experimentally that the degree of classical entanglement determines the diffraction-free propagation distance of ST wave packets. Reduction in the degree of classical entanglement manifests itself in an increased \textit{uncertainty} in the measured spatio-temporal spectral correlations. 	
\end{abstract}

\maketitle


\noindent 
Propagation-invariant wave packets are pulsed beams that are diffraction-free and dispersion-free in free space \cite{Turunen10PO,FigueroaBook14}. Underlying their wide variety is a common feature: their spatial and temporal frequencies are tightly correlated \cite{Longhi04OE,Saari04PRE}, and we thus denote them `space-time' (ST) wave packets \cite{Kondakci16OE,Parker16OE}. An idealized delta-function correlation between the spatial and temporal frequencies implies an infinite propagation distance but concomitantly an infinite energy, whereas finite-energy realizations have a finite propagation distance exceeding the usual Rayleigh range over which the beam size is relatively stable. Such wave packets have been previously realized via nonlinear phenomena \cite{DiTrapani03PRL,Faccio07OE}, approaches traditionally exploited in producing Bessel beams (e.g., axicons or annular apertures) \cite{Saari97PRL,Alexeev02PRL}, and spatio-temporal filtering \cite{Dallaire09OE,Jedrkiewicz13OE}. We have recently introduced an experimental strategy that enables precise synthesis of ST wave packets in the form of light sheets via a phase-only spatio-temporal modulation scheme that encodes arbitrary spatio-temporal spectral correlations into the field \cite{Kondakci17NP}. Exploiting this approach, we have demonstrated non-accelerating ST Airy wave packets \cite{Kondakci18PRL}, extended propagation distances \cite{Bhaduri18OE}, self-healing \cite{Kondakci18OL}, broadband ST wave packets via refractive phase plates \cite{Kondakci18OE}, and arbitrary control over the group velocity in free space \cite{Kondakci18unpub}.

Because the unique characteristics of ST wave packets stem from tight spatio-temporal spectral correlations, they are an embodiment of so-called `classical entanglement' \cite{Qian11OL,Kagalwala13NP}. In analogy to \textit{quantum} entanglement characterizing multi-partite quantum states that cannot be factored into the sub-Hilbert spaces associated with each particle, \textit{classical} entanglement is a feature of optical fields that cannot be factorized with respect to their degrees of freedom (DoFs). To date, most work on classical entanglement has focused on discretized DoFs, such as polarization and spatial modes \cite{Kagalwala13NP,Aiello15NJP}, whereas studies of classical entanglement between continuous DoFs have been lacking. 

Here we show that the propagation-invariant distance of a ST wave packet is related to the degree of classical entanglement as quantified by the Schmidt number of the field's spatio-temporal profile. We show that the degree of classical entanglement is determined by the `spectral uncertainty': the unavoidable `fuzziness' in the association between the spatial and temporal frequencies underlying the wave packet, which renders its energy and propagation distance finite \cite{Kondakci16OE}. We confirm experimentally these findings by synthesizing ST wave packets with controllable spectral uncertainty and measuring the propagation-invariant distance as the degree of classical entanglement is varied.

In quantum mechanics, starting from a \textit{pure} entangled multi-partite state, tracing out the other particles results in a \textit{mixed} single-particle state \cite{PeresBook}. For example, a two-photon entangled state displays high two-photon interference visibility, but no single-photon interference can be observed \cite{Jaeger93PRA,Abouraddy01PRA}. We consider here the analogous phenomenon occurring in classical optical fields when multiple DoFs are considered. Specifically, we examine the spatial coordinate $x$ and time $t$ of a scalar optical field $E(x,t)$ that is uniform along $y$; i.e., a light sheet. The analogue to quantum entanglement is the fact that $E(x,t)$ might not be a separable product of functions of $x$ and $t$.

A simple measure of non-separability or classical entanglement can be devised for two DoFs by assessing the lack of coherence in one DoF after tracing out the other. By expressing the field as a Schmidt decomposition \cite{Ekert95AJP,Law00PRL}, that is, a weighted sum of separable products $E(x,t)\!=\!\sum_{n}\sqrt{c_{n}}\,g_{n}(x)h_{n}(t)$, and tracing out the temporal DoF, the \textit{mutual intensity} of the field is 
\begin{equation}
J(x_{1},x_{2})=\int\!dt\,E(x_{1},t)E^{*}(x_{2},t)=\sum_{n}c_{n}\,g_{n}(x_{1})g_{n}^{*}(x_{2}),
\end{equation}
and the time-averaged intensity is $I(x)\!=\!J(x,x)\!=\!\sum_{n}c_{n}|g_{n}(x)|^{2}$. Here we assume $\int\!\!dt\,h_{n}^{*}(t)h_{m}(t)\!=\!\delta_{nm}$ and $c_{n}\!\ge\!0$ without loss of generality. The so-called spatial \textit{coherent modes} $g_{n}$ are also orthonormal, $\int\!dx\,g_{n}^{*}(x)g_{m}(x)\!=\!\delta_{nm}$. The square of the \textit{overall degree of coherence} \cite{Bastiaans84JOSA,Alonso14NJP}, which is analogous to the purity in quantum mechanics \cite{PeresBook}, is then defined as
\begin{equation}
\mu^{2}=\frac{\iint dx_{1}dx_{2}|J(x_{1},x_{2})|^{2}}{\left[\int dxI(x)\right]^2}=\frac{\sum_{n}c_{n}^{2}}{\left(\sum_{n}c_{n}\right)^2}.
\end{equation}
A simple interpretation of this measure follows from considering the case where there is only $N$ non-zero equal-magnitude coefficients $c_{n}$, whereupon $\mu^{2}\!=\!1/N$. That is, the Schmidt number $N\!=\!1/\mu^{2}$ gives the effective number of coherent modes involved. Here, $\mu^{2}\!=\!1$ denotes complete separability with one coherent mode in the Schmidt decomposition, while $\mu^{2}\!=\!0$ corresponds to maximally entangled DoFs and thus complete incoherence upon tracing out one of the DoFs. Therefore, the overall coherence of the field after time-averaging determines how non-separable the spatial and temporal DoFs are. In this Letter we consider propagating fields that depend also on the longitudinal spatial coordinate $z$. However, we focus on `propagation-invariant' fields that are independent of this DoF or nearly independent.


Consider the plane-wave expansion of a generic scalar field $E(x,z,t)$ propagating from negative to positive $z$ in free space,
\begin{equation}
E(x,z,t)=\iint\!dk_{x}d\omega\,\tilde{E}(k_{x},\omega)\,e^{i\{k_{x}x+k_{z}(k_{x},\omega)z-\omega t\}},
\end{equation}
where $k_{x}$ and $k_{z}$ are the transverse and longitudinal components of the wave vector, $\omega$ is the temporal frequency, and $k_{z}^{2}(k_{x},\omega)\!=\!(\omega/c)^{2}-k_{x}^{2}$. It is easy to show that at $z\!=\!0$ we have
\begin{equation}\label{Eq:Mu2ForField}
\mu^{2}=\frac{\int\!\!\!\int\!\!dk_{x}dk_{x}'\,|\widetilde{H}(k_{x},k_{x}')|^{2}}{\left|\int\!dk_{x}\widetilde{H}(k_{x},k_{x})\right|^{2}},
\end{equation}
where $\widetilde{H}(k_{x},k_{x}')\!=\!\int\!d\omega\widetilde{E}(k_{x},\omega)\widetilde{E}^{*}(k_{x}',\omega)$, and we normalize the spectrum $\iint\!dk_{x}d\omega|\widetilde{E}(k_{x},\omega)|^{2}\!=\!1$ such that the denominator in Eq.~\ref{Eq:Mu2ForField} is unity. We are interested in cases where the spatial frequencies $k_{x}$ are tightly correlated to the temporal frequencies $\omega$. Motivated by the realistic spectra synthesized in \cite{Kondakci17NP}, we introduce the following decomposition of the spectrum $\widetilde{E}(k_{x},\omega)\!\rightarrow\!\widetilde{E}(k_{x})\widetilde{g}(\omega-\Omega(k_{x}))$, where $\Omega(k_{x})\!=\!c\sqrt{k_{x}^{2}+k_{\mathrm{o}}^{2}}$, $c$ is the speed of light in vacuum, $\omega_{\mathrm{o}}$ is the optical carrier frequency, $k_{\mathrm{o}}\!=\!\tfrac{\omega_{\mathrm{o}}}{c}$, and $\widetilde{g}(\cdot)$ is a narrow spectral function of width $\delta\omega$ and normalized such that $\int\!d\omega|\widetilde{g}(\omega)|^{2}\!=\!1$. In the ideal limit of perfect correlation we have $\widetilde{g}(\omega)\!\rightarrow\!\delta(\omega)$. This limit corresponds to the special case of an ideal propagation-invariant field where $k_{z}$ is held constant $k_{z}\!=\!k_{\mathrm{o}}$ (Refs. \cite{Kondakci16OE,Parker16OE}), which introduces entanglement between the spatial and temporal DoFs, whereupon $E(x,z,t)\!=\!e^{ik_{\mathrm{o}}z}\!\int\!dk_{x}\widetilde{E}(k_{x})\,e^{i\{k_{x}x-\Omega(k_{x})t\}}$. For any $z$, the mutual intensity and the intensity are each the sum of two contributions,
\begin{eqnarray}
J(x_{1},x_{2})\!\!\!&=&\!\!\!\!\!\int\!dk_{x}|\widetilde{E}(k_x)|^2\sqrt{k_{\mathrm{o}}^{2}/k_{x}^{2}+1}\,\,e^{ik_{x}(x_{2}-x_{1})}\nonumber\\
\!\!\!&+&\!\!\!\!\!\Re\left\{\!\!\int\!dk_{x}\widetilde{E}(k_{x})\widetilde{E}^{*}(-k_{x})\!\sqrt{\!k_{\mathrm{o}}^{2}\!/k_{x}^{2}\!+\!1}\,e^{ik_{x}(x_{1}+x_{2})}\!\right\}\!,\\
I(x)\!\!\!&=&\!\!\!\!\!\int_0^{\infty}\!dk_{x}I_{0}(k_{x})\left\{1+\eta(k_{x})\cos[2k_{x}x-\phi(k_{x})]\right\},
\end{eqnarray}
where
\begin{eqnarray}
I_{0}(k_{x})\!\!\!\!&=&\!\!\!\sqrt{k_{\mathrm{o}}^{2}\!/k_{x}^{2}\!+\!1}\,\left[|\widetilde{E}(k_{x})|^{2}+|\widetilde{E}(-k_{x})|^2\right],\\ 
\eta(k_{x})\!\!\!\!&=&\!\!\!2|\widetilde{E}(k_{x})\widetilde{E}^*(-k_{x})|/\left(|\widetilde{E}(k_{x})|^{2}+|\widetilde{E}(-k_{x})|^{2}\right),\\
\phi(k_{x})\!\!\!\!&=&\!\!\!-{\rm arg}[\widetilde{E}(k_{x})\widetilde{E}^*(-k_{x})].
\end{eqnarray}
Because $0\!\le\!\eta\!\le\!1$, it is clear that the transverse intensity profile is composed of a uniform pedestal $\int\!dk_{x}I_{0}(k_{x})$ plus a continuous superposition of sinusoidals that can be used to construct any desired functional form through standard Fourier theory on top of that pedestal, with the constraint that the magnitude of the largest feature of the constructed function is not larger than the height of the pedestal. This result is easy to understand: for each temporal frequency we have the superposition of two plane waves with spatial frequencies $\pm k_x$, whose intensity is precisely the non-negative sum of a constant and a sinusoidal. Since the interference between different temporal frequency components is erased by the time average, we have the superposition of the corresponding spatial intensity distributions. 

Calculating the measure $\mu^{2}$ for this idealized case is involved, and is best done by first evaluating the spatial integrals and then the temporal average. One finds that $\mu^{2}\!=\!0$; i.e., the Schmidt number is infinite and the degree of entanglement is maximal. This is consistent with the fact that the field is an idealization with infinite extent. Note that this result is \textit{independent} of the particular form of $\widetilde{E}(k_{x})$ and depends solely on the fact that a perfect (delta-function) correlation exists between $k_{x}$ and $\omega$.

\begin{figure}[t!]
	\begin{center}
		\includegraphics[scale=0.9]{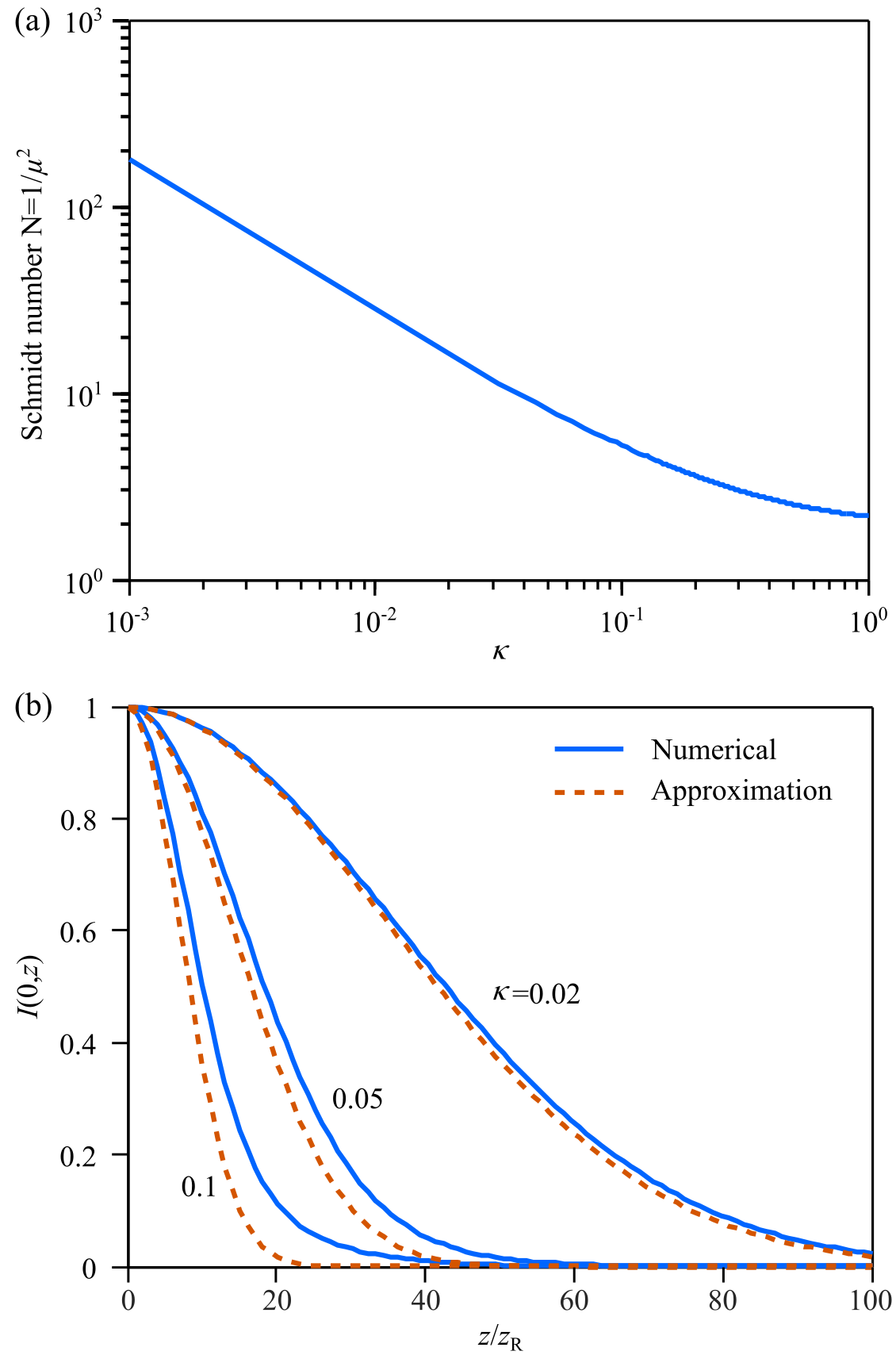}\vspace*{-3mm}
	\end{center}
	\caption{(a) Calculated Schmidt number $N\!=\!1/\mu^{2}$ as a function of $\kappa$. (b) Calculated $I(0,z)$ and the approximation $I(0,z)\!\approx\!\exp{\{-(\kappa\tfrac{z}{z_{\mathrm{R}}})^{2}\}}$. The intensity drops by half at $z\!\!\approx\!z_{\mathrm{R}}/\kappa$; i.e., classical entanglement -- which is related monotonically to $\kappa$ -- increases the propagation distance.}
	\label{Fig:Simulations}
\end{figure}

We now relax the requirement of perfect delta-function correlation between $k_{x}$ and $\omega$ and introduce an uncertainty in their association, and take the simple form of a Gaussian function for the spectral uncertainty $\widetilde{g}(\omega)\!\propto\!\exp{\{-\tfrac{1}{2}(\tfrac{\omega-\omega_{\mathrm{o}}}{\delta\omega})^{2}\}}$ whereupon
\begin{equation}
\mu^{2}\!=\!\!\int\!\!\!\!\int\!\!dk_{x}dk_{x}'|\widetilde{E}(k_{x})|^{2}|\widetilde{E}(k_{x}')|^{2}\exp{\!\left\{-\frac{[\Omega(k_{x})\!-\!\Omega(k_{x}')]^{2}}{2(\delta\omega)^{2}}\right\}}.
\end{equation}
Taking a Gaussian spatial spectrum $\widetilde{E}(k_{x})\!\propto\!\exp{\{-k_{x}^{2}/(2\Delta_{x}^{2})\}}$ and making the approximation $\Omega(k_{x})-\omega_{\mathrm{o}}\!\approx\!k_{x}^{2}/(2k_{\mathrm{o}}^{2})$,
\begin{equation}
\mu^{2}\!=\!\frac{1}{\pi\Delta_{x}^{2}}\int\!\!\!\int\!\!dk_{x}dk_{x}'\exp{\left\{-\frac{k_{x}^{2}+k_{x}'^{2}}{2\Delta_{x}^{2}}\right\}}\exp{\left\{-\frac{(k_{x}^{2}-k_{x}'^{2})^{2}}{8k_{\mathrm{o}}^{4}(\delta\omega)^{2}/\omega_{\mathrm{o}}^{2}}\right\}},
\end{equation}
and simplifying the integrals yields the expression
\begin{equation}
\mu^{2}=\frac{1}{2\sqrt{\pi}}\int_{0}^{\infty}
\!\!\!\frac{e^{-t}}{\sqrt{t(1+\frac{t}{2\kappa^{2}}})}dt=\frac{1}{\sqrt{2\pi}}K_{0}(\kappa^{2})e^{\kappa^{2}}\kappa,
\end{equation}
where $K_{0}$ is the 0$^{\mathrm{th}}$-order modified Bessel function of the second kind, $\kappa\!=\!\tfrac{\delta\omega/\omega_{\mathrm{o}}}{\Delta_{x}^{2}/k_{\mathrm{o}}^{2}}\!=\!\tfrac{\delta\omega}{\Delta\omega}$ is the ratio of the spectral uncertainty $\delta\omega$ to the full spectral bandwidth $\Delta\omega$, and normally $\kappa\!\ll\!1$. As shown in Fig.~\ref{Fig:Simulations}(a), the Schmidt number $N\!=\!1/\mu^{2}$ is now finite and drops with increasing $\kappa$, indicating that \textit{the degree of classical entanglement of the field increases with reduced uncertainty}.

Concomitantly with the drop in the degree of classical entanglement, the field now features $z$-dependent axial dynamics. To examine these dynamics, we evaluate the time-averaged intensity $I(x,z)\!=\!\int\!dt|E(x,z,t)|^{2}$ after making use of the Gaussian forms for $\widetilde{E}(k_{x})$ and $\widetilde{g}(\omega)$ utilized above, whereupon
\begin{equation}\label{Eq:FullTimeAveragedIntensity}
I(x,z)=\Delta_{x}\int_{0}^{\infty}\!dt\frac{e^{-t}}{\sqrt{t(1+\frac{t}{\kappa^{2}})}}\exp{\left\{-\frac{(\sqrt{t}\frac{z}{z_{\mathrm{R}}}-x\Delta_{x})^{2}}{1+\frac{t}{\kappa^{2}}}\right\}},
\end{equation}
where $z_{\mathrm{R}}\!=\!\tfrac{k_{\mathrm{o}}}{\Delta_{x}^{2}}$ is the Rayleigh range of a Gaussian beam of spatial bandwidth $\Delta_{x}$ and frequency $\omega_{\mathrm{o}}$. For small spectral uncertainty $\kappa\!\ll\!1$, the field on axis is $I(0,z)\widetilde{\propto}\exp{\{-(\kappa\frac{z}{z_{\mathrm{R}}})^{2}\}}$, and the approximation improves for smaller $\kappa$ [Fig.~\ref{Fig:Simulations}(b)]. That is, the Rayleigh range is extended by a factor equal to $1/\kappa$, which is related monotonically to the Schmidt number, demonstrating that the degree of classical entanglement dictates the propagation distance of this ST wave packet. Furthermore, Eq.~\ref{Eq:FullTimeAveragedIntensity} indicates that the pedestal now has a finite width that is related to the propagation distance. 

\begin{figure}[t!]
	\begin{center}
		\includegraphics[width=9.2cm]{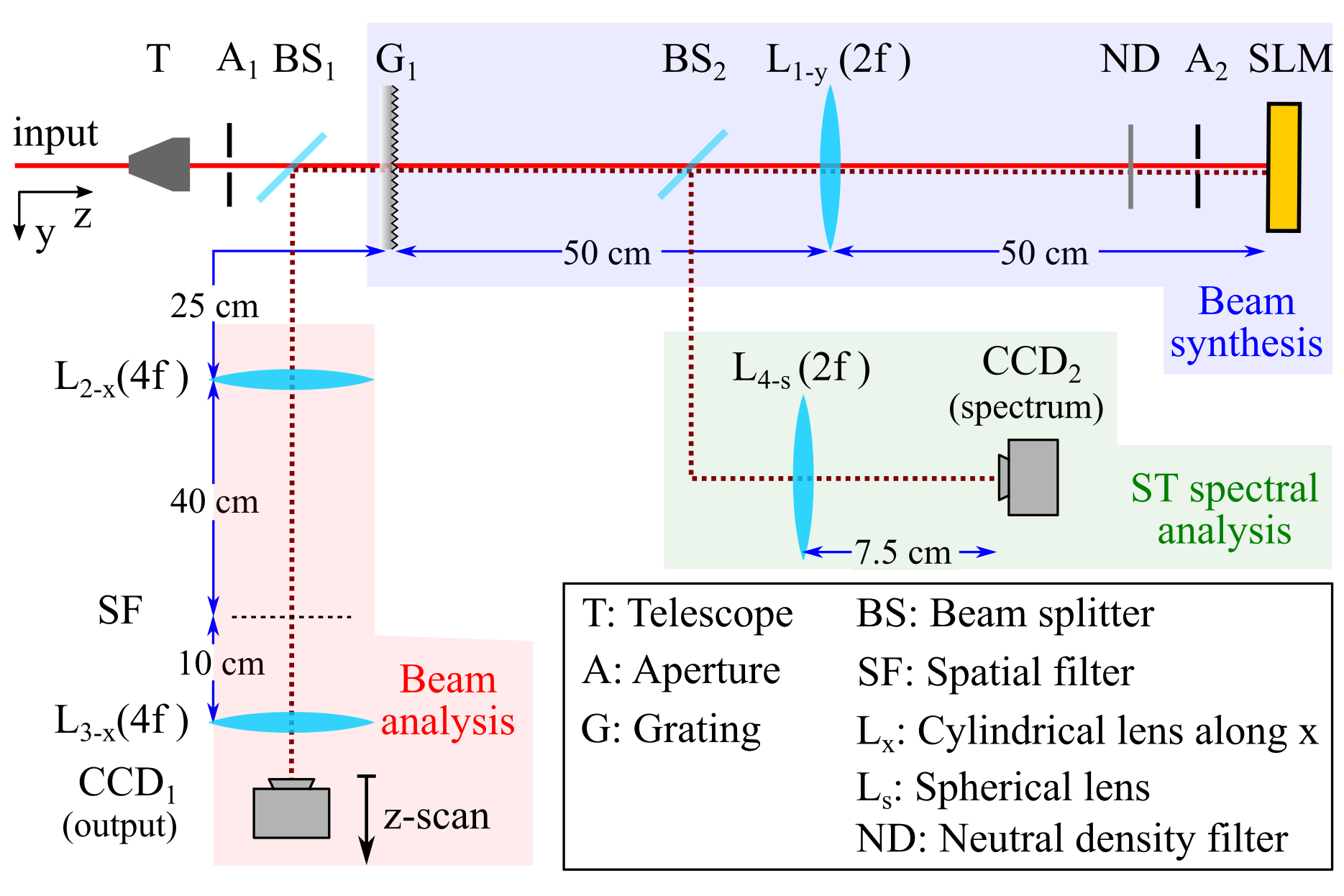}\vspace*{-3mm}
	\end{center}
	\caption{Schematic depiction of the optical setup for synthesis and analysis of a ST wave packet. The inset provides a key for the optical components. The various sections of the setup are identified with differently colored backgrounds.}
	\label{Fig:Setup}
\end{figure}

We now proceed to demonstrating experimentally that the spectral uncertainty $\delta\omega$, and thus the degree of classical entanglement, determines the propagation-invariant distance. The setup for synthesizing the ST wave packets is shown in Fig.~\ref{Fig:Setup}. The spectrum of pulses from a femtosecond Ti:Sa laser (Tsunami, Spectra Physics; central wavelength $\sim\!800$~nm) is spread spatially with a diffraction grating (1200~lines/mm and area $25\times25$~cm$^2$) and collimated by a cylindrical lens before impinging on a spatial light modulator (SLM; Hamamatsu X10468-02) that imparts a phase distribution to associate each pair of spatial frequencies $\pm k_{x}$ with a single wavelength $\lambda$. The phase distribution is designed to produce the particular ST wave packets for which $k_{z}\!=\!k_{\mathrm{o}}$. The retro-reflected field from the SLM passes through the cylindrical lens back to the grating, whereupon the ST wave packet is formed as the pulse is reconstituted. The ST wave packet is characterized in the spectral domain where we obtain the spatio-temporal spectrum $|\tilde{E}(k_{x},\lambda)|^{2}$ by implementing a spatial Fourier transform to the spread spectrum, and in physical space where we record the time-averaged intensity $I(x,z)\!=\!|\int\!dt\,E(x,z,t)|^{2}$.

\begin{figure*}[t!]
	\begin{center}
		\includegraphics[width=18.6cm]{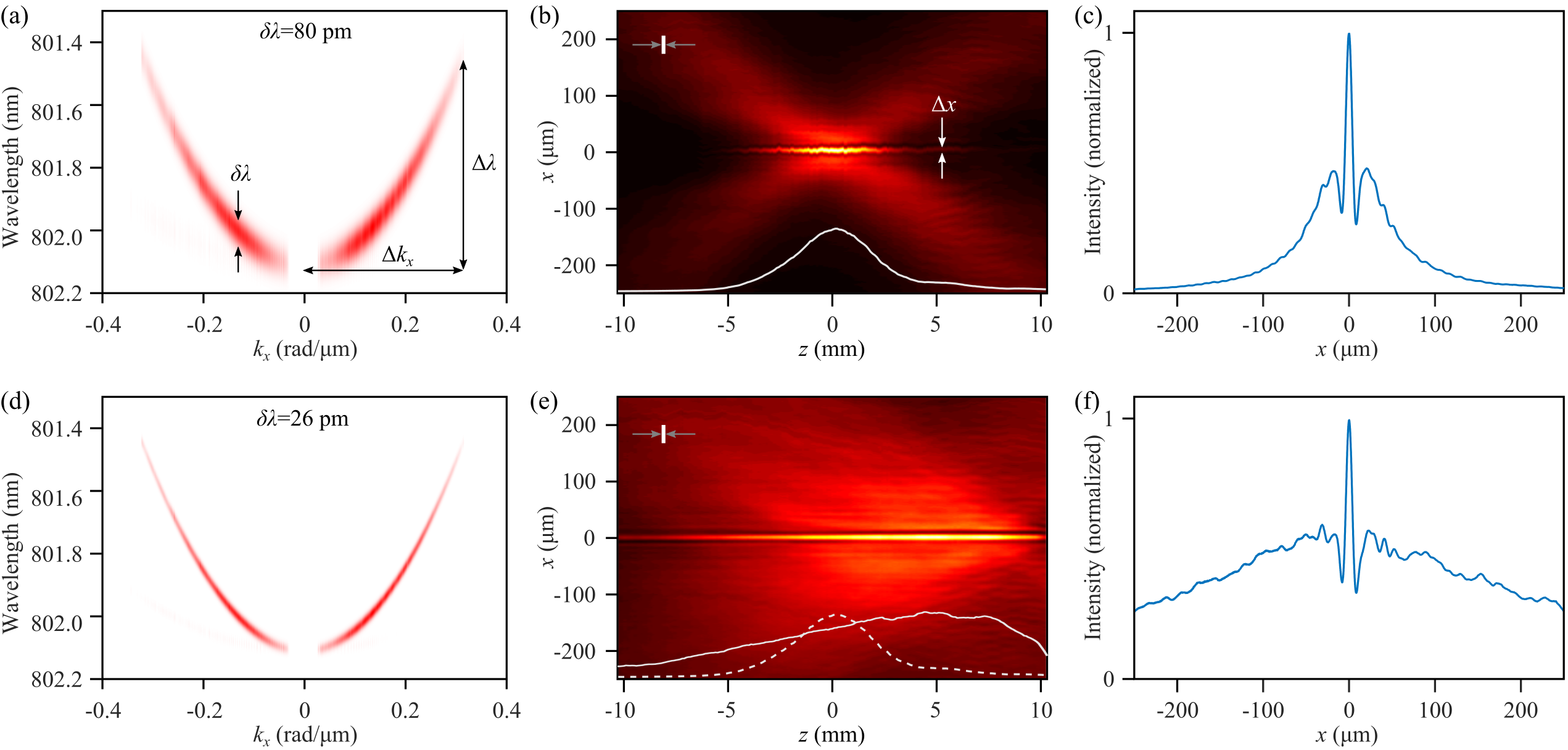}\vspace*{-3mm}
	\end{center}
	\caption{Measurements of the dependence of the propagation-invariance distance on the spectral uncertainty $\delta\lambda$. (a) Measured spatio-temporal spectrum $|\tilde{E}(k_{x},\lambda)|^{2}$ for $\delta\lambda\!=\!80$~pm. (b) Measured axial propagation of the time-averaged intensity $I(x,z)$. The white curve is the axial intensity $I(0,z)$. (c) Axially averaged intensity $I(x)\!=\!\int\!dz\,I(x,z)$ from (b). (d)-(f) Same as (a)-(c), with $\delta\lambda\!=\!26$~pm.}
	\label{Fig:Spectra}
\end{figure*}

\begin{figure}[h!]
	\begin{center}
		\includegraphics[width=9.2cm]{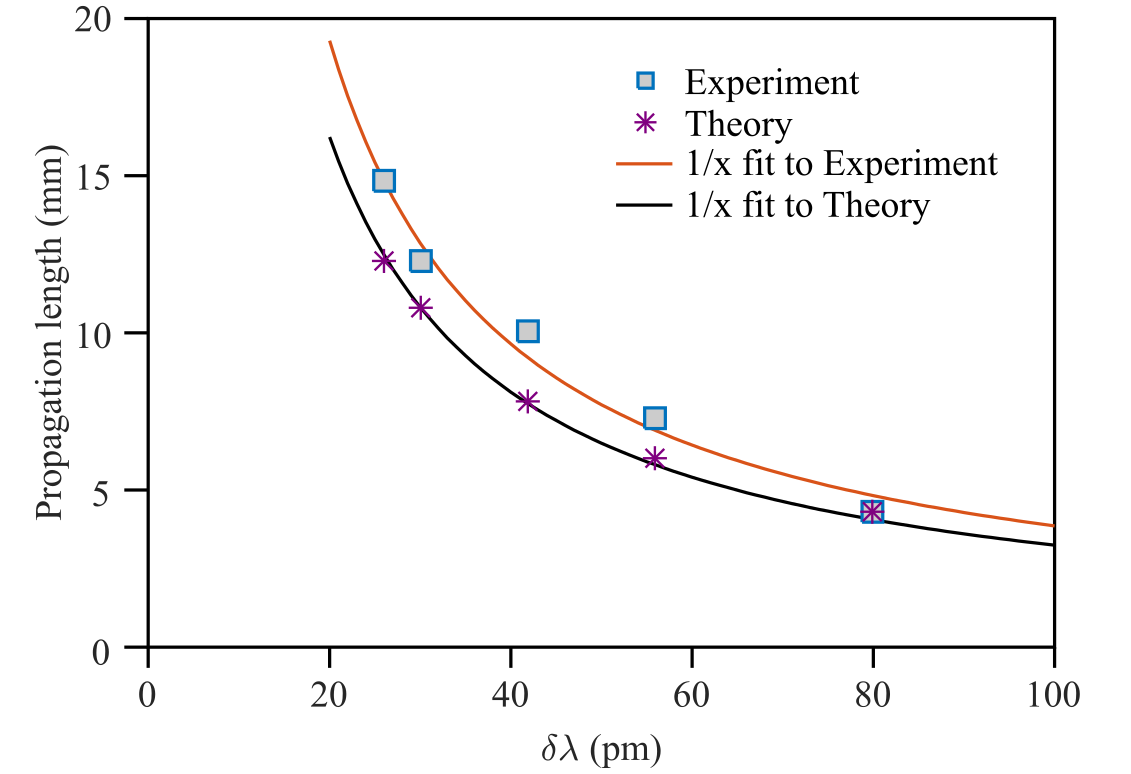}\vspace*{-3mm}
	\end{center}
	\caption{Measured and calculated propagation-invariant distance while varying the spectral uncertainty $\delta\lambda$ for fixed temporal and spatial bandwidths, $\Delta\lambda$ and $\Delta k_{x}$, respectively. For small values of spectral uncertainty, $\delta\lambda\widetilde{\propto}\kappa$.}
	\label{Fig:Data}
\end{figure}

In all cases, the temporal bandwidth is $\Delta\lambda\!\sim\!0.7$~nm and the spatial bandwidth is $\Delta k_{x}\!\sim\!0.3$~rad/$\mu$m (half-width at half-maximum), corresponding to a transverse width of the spatial profile of $\Delta x\!\approx\!8$~$\mu$m. The spectral uncertainty $\delta\lambda$ is varied in the range 20 to 80~pm by limiting the illuminated area on the diffraction grating from 25~mm to 5~mm to change the grating's spectral resolving power. Two examples of measured ST wave packets are presented in Fig.~\ref{Fig:Spectra}. Data for an ST wave packet having a spectral uncertainty of $\delta\lambda\!=\!80$~pm [Fig.~\ref{Fig:Spectra}(a)-(c)] shows that the spatio-temporal spectrum $|\tilde{E}(k_{x},\lambda)|^{2}$ conforms to theory [Fig.~\ref{Fig:Spectra}(a)], the axial evolution of the intensity $I(x,z)$ [Fig.~\ref{Fig:Spectra}(b)] reveals a propagation-invariant distance of $\sim5$~mm and a narrow pedestal [Fig.~\ref{Fig:Spectra}(c)]. Reducing $\delta\lambda$ to $26$~pm [Fig.~\ref{Fig:Spectra}(d)] results in an increase in the propagation distance to $\sim15$~mm [Fig.~\ref{Fig:Spectra}(e)] \textit{without} changing $\Delta x$, and a concomitant increase in the width of the pedestal [Fig.~\ref{Fig:Spectra}(f)]. Measured values of the propagation distance while varying $\delta\lambda$ -- but holding $\Delta\lambda$ and $\Delta k_{x}$ fixed -- are plotted in Fig.~\ref{Fig:Data}. The data points fit a $\tfrac{1}{x}$-relationship as expected from the theoretical $\tfrac{1}{\kappa}$-dependence of the propagation distance.

The framework that we have introduced can be extended to other ST wave packets beyond those having a fixed axial wave vector component $k_{z}\!=\!k_{\mathrm{o}}$. Indeed, rigid transport of the field envelope $|E(x,z,t)|\!=\!|E(x,0,t-z/v_{\mathrm{g}})|$ implies that $\omega-\omega_{\mathrm{o}}\!=\!(k_{z}-k_{\mathrm{o}})v_{\mathrm{g}}$, where $v_{\mathrm{g}}$ is the group velocity. This constraint indicates that $k_{x}$ and $\omega$ are related through the equation of a conic section \cite{Kondakci17NP}. It can be shown that the linear correlation between $k_{z}$ and $\omega$ entails that $\mu^{2}\!=\!0$, indicating perfect entanglement. A full classification of ST wave packets is provided in \cite{Yessenov18PRA}, and it is important to study the impact of classical entanglement on the propagation-invariance of these various classes.

In conclusion, we have shown theoretically that the degree of classical entanglement -- quantified by the Schmidt number of the field -- determines the propagation-invariant distance of ST wave packets. The time-averaged intensity of ST wave packets comprises a narrow spatial feature atop a broad pedestal. Reduction in classical entanglement manifests itself in an increase in the uncertainty of the field spatio-temporal spectral correlations, and is accompanied by a decrease in the propagation distance and a narrowing of the pedestal without changing the transverse beam width atop the pedestal. We have verified these predictions experimentally by synthesizing ST wave packets with controllable spectral uncertainty.\\

\noindent \textbf{Funding}\\
U.S. Office of Naval Research (ONR) (N00014-17-1-2458) for HEK and AFA. National Science Foundation (NSF) (PHY-1507278); the Excellence Initiative of Aix-Marseille University -- A*MIDEX, a French ``Investissements d’Avenir'' programme for MAA.

\newpage 
\bibliography{diffraction}

\end{document}